\documentclass[12pt]{iopart}
\usepackage[utf8]{inputenc}
\usepackage[colorlinks=true,linkcolor=blue,citecolor=blue,urlcolor=blue]{hyperref}
\usepackage{svg}
\usepackage[T1]{fontenc}
\usepackage{bm}
\usepackage{subcaption}
\usepackage{multirow}
\usepackage{float}
\usepackage{setspace}
\usepackage{graphicx}
\usepackage{siunitx}
\DeclareSIUnit\rydberg{Ry}
\usepackage{booktabs}
\usepackage[singlelinecheck=false]{caption}
\usepackage{balance} 
\usepackage{cite}
\usepackage{lineno}

\begin{document}

\title[Noncollinear magnetism in a monolayer of 2D CrTe$_2$]{Noncollinear magnetism in two-dimensional CrTe$_2$}
\author{Nihad Abuawwad$^{1,2,3}$, Manuel dos Santos Dias$^{2,1,4}$, Hazem Abusara$^{3}$, Samir Lounis$^{1,2}$}
\address{$^1$ Peter Gr\"unberg Institut and Institute for Advanced Simulation, Forschungszentrum J\"ulich \& JARA, 52425 J\"ulich, Germany}
\address{$^2$ Faculty of Physics, University of Duisburg-Essen, 47053 Duisburg, Germany}
\address{$^3$ Department of Physics, Birzeit University, PO Box 14, Birzeit, Palestine}
\address{$^4$ Scientific Computing Department, STFC Daresbury Laboratory, Warrington WA4 4AD, United Kingdom}
\eads{\mailto{n.abuawwad@fz-juelich.de}, \mailto{s.lounis@fz-juelich.de}}

\begin{abstract}
The discovery of two-dimensional (2D) van der Waals magnets opened unprecedented opportunities for the fundamental exploration of magnetism in quantum materials and the realization of next generation spintronic devices.
Here, based on a multiscale modelling approach that combines first-principles calculations and a Heisenberg model supplied with ab-initio parameters, we report a strong magnetoelastic coupling in a free-standing monolayer of CrTe$_2$.
We demonstrate that different crystal structures of a single CrTe$_2$ give rise to non-collinear magnetism through magnetic frustration and emergence of the Dzyaloshinskii-Moriya interaction (DMI).
Utilizing atomistic spin dynamics, we perform a detailed investigation of the complex magnetic properties pertaining to this 2D material impacted by the presence of various types of structural distortions akin to charge density waves.
\end{abstract}
\submitto{\JPCM}

\maketitle

\section{Introduction}
Back in 1966, Mermin and Wagner proved that two-dimensional (2D) systems with continuous symmetry cannot be magnetically ordered at finite temperature\cite{mermin}.
This makes magnetic order in 2D reliant on interactions that breaks the spin-rotational invariance, for instance, an external magnetic field or the magnetocrystalline anisotropy, which are typically much weaker than exchange interactions and so might lead to very low ordering temperatures.
It was thus very surprising that 2D magnets were discovered experimentally in 2017 in CrI$_{3}$ \cite{cri3} and Cr$_{2}$Ge$_{2}$Te$_{6}$ \cite{cr2ge2te2} down to the mono- and bilayer limit, respectively.
Afterwards, intense research activities were made to expand the development of 2D magnets \cite{Gibertini2019, Gong2019, McGuire2020, Jiang2021b, Sierra2021}.
In most cases, these 2D materials have a simple collinear magnetic order, such as ferromagnetic for VI$_{3}$, CrCl$_{3}$, and CrBr$_{3}$, or antiferromagnetic for NiPS$_{3}$, FePS$_{3}$, and MnPS$_{3}$, but they could also display complex magnetism, for example by hosting skyrmions and other non-collinear spin textures \cite{WTe2}.

CrTe$_2$ is a new entrant in the field of 2D magnets, although it has been known for a few years in bulk form \cite{Freitas2015} and having been simulated in monolayer form \cite{Lv2015}.
Recently, magnetic circular dichroism experiments demonstrated that thin CrTe$_2$ grown either on SiO$_2$/Si or bilayer graphene substrates are ferromagnetic with a Curie temperature of \SI{200}{\kelvin} \cite{fm-crte2,Zhang2021}.
This Curie temperature is  high when compared to other 2D magnetic materials such as CrI$_{3}$, which as a single monolayer has a Curie temperature of \SI{45}{\kelvin}.
In contrast to the previous finding, a monolayer of CrTe$_2$ deposited on graphene was found to host a zig-zag antiferromagnetic state as revealed by spin-polarised scanning tunnelling microscopy (SP-STM) combined with first-principles calculations, with an applied magnetic field driving the monolayer into a noncollinear spin texture \cite{afm-crte2}. 
Interestingly, when simulating a free standing CrTe$_2$  by density functional theory (DFT), a charge density wave (CDW) phase has been found after observing a clear instability in the phonon band structure \cite{cdw-crte2}.  
Further theoretical studies unveiled the strong dependence of the magnetic ground states of 1T-CrTe$_2$ on the film thickness: an intralayer and interlayer antiferromagnetic-ferromagnetic transition occurs at a critical thickness
of five CrTe$_2$ layers, which represents the bulk magnetic state \cite{thick-crte2}.
Moreover, DFT calculations for monolayer of CrTe$_{2}$ show an antiferromagnetic (AFM) metallic behaviour in 1T phase, and ferromagnetic (FM) semiconductor in deformed phase of 1T called 1T$^{\prime}$ phase \cite{diverse-crte2}.

In the present study, we rationalize the diverse behavior of CrTe$_2$ by demonstrating using first-principles calculations a strong coupling between magnetism and crystal structure in a single layer of CrTe$_{2}$, whose magnetic states are subsequently explored via atomistic spin dynamics.

\section{Computational methods}

\subsection{First-principles calculations}
Atomic relaxations as function of various collinear magnetic states of CrTe$_{2}$ were assessed using DFT as implemented in the Quantum Espresso (QE) computational package \cite{qe} with projector augmented plane wave (PAW) pseudopotentials \cite{ps}.
In our calculations, the generalized gradient approximation (GGA) of Perdew-Burke-Ernzerhof (PBE)\cite{PhysRevLett.77.3865} was used as the exchange and correlation functional.
The plane-wave energy cut-off is \SI{80}{\rydberg}, and the convergence criterion for the total energy is set to \SI{0.1}{\micro\electronvolt}.
We included a vacuum region of \SI{20}{\angstrom} in the direction normal to the plane of the monolayer to minimize the interaction between the periodic images. 
The residual forces on the relaxed atomic positions  were smaller than \SI{0.01}{\electronvolt\per\angstrom}, and the strain on the unit cell smaller than \SI{0.5}{\kilo\bar}. 
The self-consistent calculations were performed with a k-mesh of $24 \times 24 \times 1$ points for the unit cell of 1T-phase.
The Brillouin zone summations used a Gaussian smearing of \SI{0.01}{\rydberg}.

Once the geometries of the various collinear magnetic states were established, we explored in detail magnetic properties and interactions with the all-electron full-potential relativistic Korringa-Kohn-Rostoker Green function (KKR-GF) method as implemented in the JuKKR computational package \cite{Papanikolaou2002,Bauer2014,jukkr}. 
The angular momentum expansion of the Green function was truncated at $\ell_\mathrm{max} = 3$ with a k-mesh of $48 \times 48 \times 1$ points.
The energy integrations were performed including a Fermi-Dirac smearing of \SI{502.78}{\kelvin}.
The Heisenberg exchange interactions and Dzyaloshinskii–Moriya (DM) vectors were extracted using the infinitesimal rotation method \cite{inf-rot} with a finer k-mesh of $200\times 200 \times 1$.

\subsection{Magnetic interactions and atomistic spin dynamics}
The magnetic interactions obtained from the first-principles calculations are used to parameterize the following classical Heisenberg hamiltonian with unit spins, $|\vec{S}| = 1$, which includes the Heisenberg exchange coupling ($J$), the DM interaction ($D$), the magnetic anisotropy energy ($K$), and the Zeeman term ($B$):
\begin{eqnarray}
\label{eq:spin_model}
    H &= - \sum_{n\mu\nu} J_{0\mu,n\nu}\vec{S}_{0\mu} \cdot \vec{S}_{n\nu} - \sum_{n\mu\nu} \vec{D}_{0\mu,n\nu}\cdot\left(\vec{S}_{0\mu} \times \vec{S}_{n\nu}\right) \nonumber\\
    &+ \sum_{n\mu} K_{n\mu} \left(S_{n\mu}^z\right)^2 -\sum_{n\mu} \vec{B}\cdot \vec{S}_{n\mu}
\end{eqnarray}
Here $n$ labels units cells with 0 being the one including the origin, and $\mu$ and $\nu$ label different magnetic sites within a unit cell.
The magnetic properties pertaining to CrTe$_{2}$ were evaluated by analysing the Fourier-transformed magnetic interactions, which in reciprocal space gives access to the magnetic ground state and the related dispersion of potential spin spirals:
\begin{equation}
\label{eq:exc_contr}
    J_{\mu\nu}(\vec{q}) = \sum_n J_{0\mu,n\nu}e^{-i\vec{q}\cdot(\vec{R}_{0n}+\vec{R}_{\mu\nu})}
\end{equation}
where $\vec{R}_{0n}$ is a vector connecting unit cells 0 and $n$, while $\vec{R}_{\mu\nu}$ is a vector connecting atoms $\mu$ and $\nu$ in the same unit cell.

Furthermore, atomistic spin dynamic simulations using the Landau-Lifshitz-equation (LLG) as implemented in the Spirit code \cite{Mueller2019a,spirit} are performed in order to validate the predictions from eq.~\ref{eq:exc_contr} and to explore whether more complex magnetic states can arise.
We used the simulated annealing method: we started from a random spin state at \SI{1000}{\kelvin} which we let equilibrate, then cool the system in steps by reducing the temperature to half of its previous value and equilibrating again, until we reach below \SI{10}{\kelvin}.

\section{Results and discussion}

\subsection{Geometric and magnetic properties}
\begin{figure}[htb]
    \begin{center}\textbf{Ferromagnetic states}\par\medskip\end{center}\vspace{-1em}
    \begin{subfigure}[b]{0.25\textwidth}
        \captionsetup{justification=centering}
        \caption{1T}
        \includegraphics[width=\textwidth]{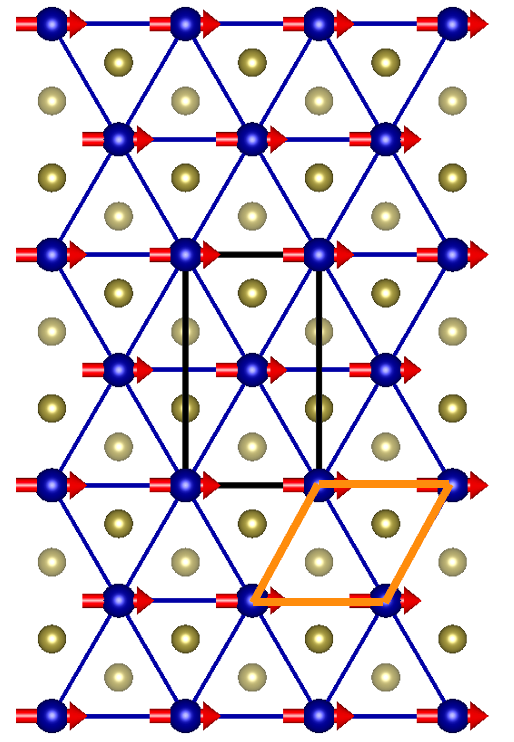}
    \end{subfigure}
    \hspace{0.12\textwidth}
    \begin{subfigure}[b]{0.24\textwidth}
        \captionsetup{justification=centering}
        \caption{1T$^{\prime}$}
        \includegraphics[width=\textwidth]{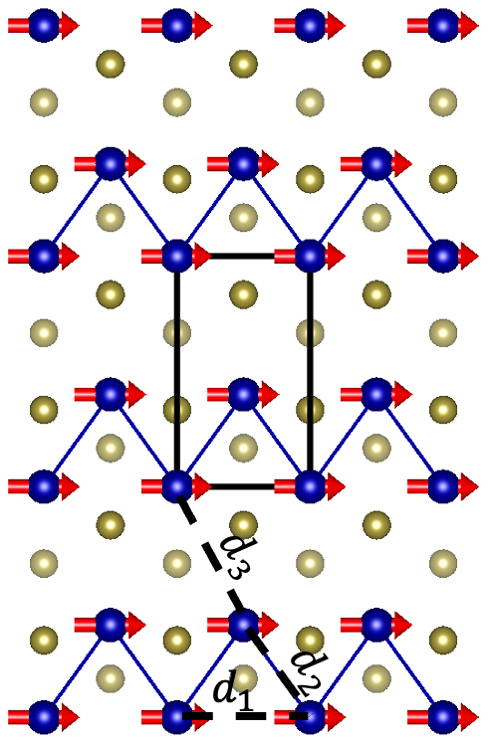}
    \end{subfigure}
    \hspace{0.12\textwidth}
    \begin{subfigure}[b]{0.24\textwidth}
        \captionsetup{justification=centering}
        \caption{CDW}
        \includegraphics[width=\textwidth]{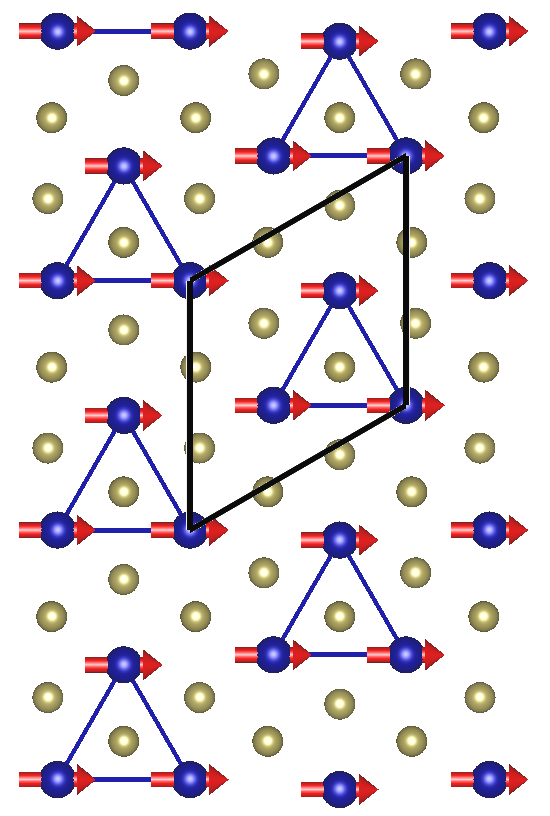}
    \end{subfigure}
    \begin{center} \par\medskip \textbf{Antiferromagnetic states}\par\medskip \end{center}\vspace{-1em}
    \begin{subfigure}[b]{0.25\textwidth}
        \captionsetup{justification=centering}
        \caption{1T$^{\prime\prime}$}
        \includegraphics[width=\textwidth]{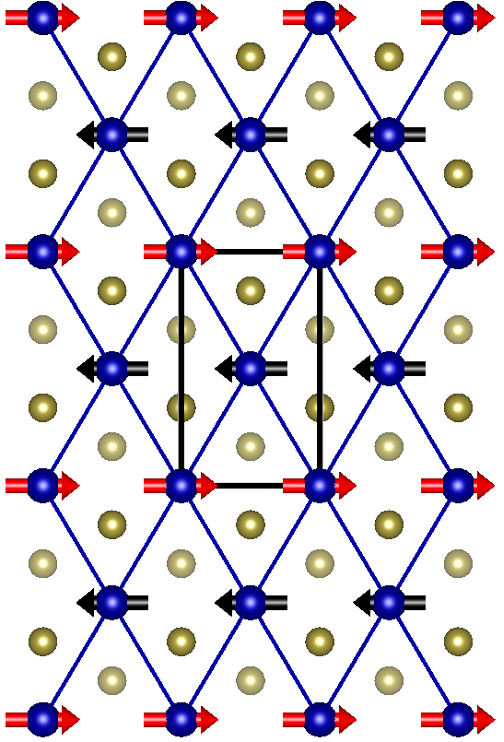}
    \end{subfigure} 
    \hspace{0.12\textwidth}
    \begin{subfigure}[b]{0.255\textwidth}
        \captionsetup{justification=centering}
        \caption{AABB}
        \includegraphics[width=\textwidth]{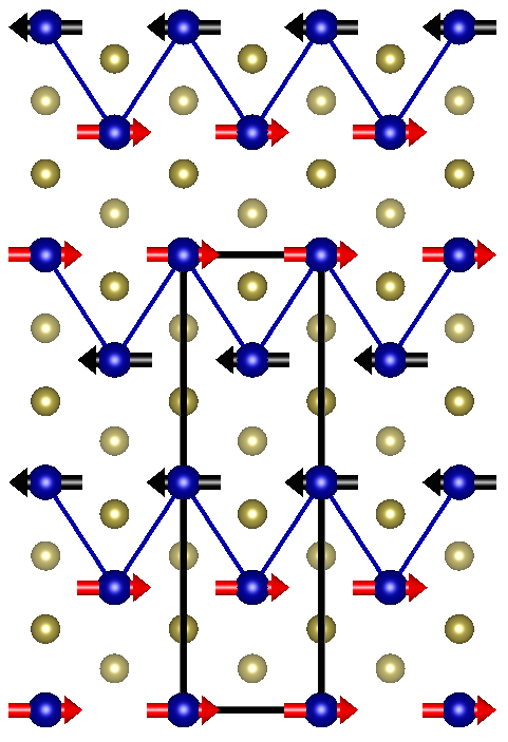}
    \end{subfigure} 
    \hspace{0.1\textwidth}
    \begin{subfigure}[b]{0.25\textwidth}
        \captionsetup{justification=centering}
        \caption{Zig-zag}
        \includegraphics[width=\textwidth]{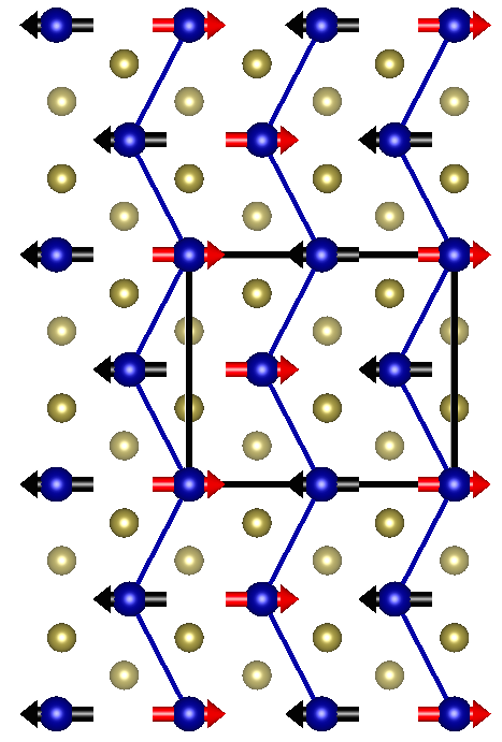}
    \end{subfigure} 
    \caption{\label{structure}
    Crystal structures of CrTe$_{2}$ monolayer for different asssumed magnetic states.
    (a-c) Crystal structures of FM phases: 1T, 1T$^{\prime}$, and CDW.
    1T$^{\prime}$ corresponds to a $1\times \sqrt{3}$ and CDW to a $\sqrt{3}\times \sqrt{3}$ supercell of 1T, respectively.
    (d-f) Crystal structures of AFM phases: 1T$^{\prime\prime}$, AABB, and zig-zag.
    Cr and Te atoms are in blue and gold colors, respectively.
    The computational unit cells for the different structures are shown by the black rectangles or rhombi, while the primitive cell for the 1T structure is indicated by the orange rhombus.
    The different orientations of the magnetic moments are represented by red and black arrows.
    Blue lines indicate the shortest bonds between the Cr atoms, providing an easy way to visually differentiate the crystal structures.}     
\end{figure}

We start with the results of our QE simulations.
Fig.\ \ref{structure}(a-c) shows the three crystal structures (1T, 1T$^{\prime}$, and CDW) found for the ferromagnetic state of a CrTe$_{2}$ monolayer, which differ in the shortest bond arrangements between the Cr atoms. 
For the 1T phase, the primitive unit cell is a 120$^{\circ}$ rhombus, while the 1T$^{\prime}$ phase can be regarded as a distorted structure from the 1T phase, arising from a Peierls-like instability; this is a primary mechanism for the formation of the 1T$^{\prime}$ phase in transition metal dichalcogenides (TMDs) such as MoS$_{2}$ \cite{mos} and WTe$_{2}$ \cite{wte}. 
The primitive cell of 1T$^{\prime}$ is a rectangular unit cell, corresponding to a $1 \times \sqrt{3}$ supercell of the 1T phase. 
Compared to the latter one, two adjacent rows of Cr atoms in the vertical direction move toward each other.
Similarly, the CDW structure can be regarded as an alternative distortion of the 1T phase, which was identified and traced to a phonon instability in Ref.\ \cite{cdw-crte2}.
The primitive cell of CDW is a 30$^{\circ}$ rotated $\sqrt{3} \times \sqrt{3}$ hexagonal cell with respect to the 1T phase. 

The bonds connecting a Cr atom to its neighbors are useful to distinguish between the unveiled structures, and so we define three distances, $d_1$, $d_2$ and $d_3$ (see Fig.\ \ref{structure}), which are equal in the 1T phase, two of them are equal in the CDW case ($d_1$ = $d_2$ $\neq$ $d_3$), and all different for the 1T$^{\prime}$ phase.
The shortest bonds can be used to visually distinguish the structures, and so are shown in blue in Fig.\ \ref{structure}.
The Cr magnetic moments are about 2.6 $\mu_\mathrm{B}$ and are listed along with the bond lengths in Table \ref{table:structure}.

When imposing an initial AFM state to either 1T or 1T$^{\prime}$ structures, the geometry changes substantially and we end up with a new phase that we coin 1T$^{\prime\prime}$, shown in Fig. \ref{structure}(d-f), where $d_1$ $\neq$ $d_2$ = $d_3$.
We also considered more complex AFM arrangements by expanding the 1T$^{\prime\prime}$ unit cell to build up the AABB structure where two rows of Cr magnetic moments align parallel (AA) followed by two rows which align anti-parallel to the first two rows (BB), or a zig-zag structure where the parallel magnetic moments are arranged in a zig-zag pattern which is followed by another zig-zag of anti-parallel magnetic moments. 
The AFM magnetic moments as listed in Table \ref{table:structure} tend to be smaller than the FM ones. 

Overall, our findings highlight the strong magneto-elastic coupling characterizing the free-standing CrTe$_2$ monolayer.
Imposing different magnetic states leads to large forces and stresses on the unit cell, from which emerge new crystal structures by energy minimization.

\begin{table}[htb]
\caption{\label{table:structure}Lattice parameters, bond lengths and spin moments of the different CrTe$_2$ monolayer structures hosting various FM and AFM phases.}
\begin{tabular*}{\textwidth}{l @{\extracolsep{\fill}} lllllll}
\br
&\centre{5}{Lattice parameters (\AA)} \\ \cmidrule{2-6} 
Phase& $a$ & $b$ & $d_1$ & $d_2$ & $d_3$ & & $m$ ($\mu_\mathrm{B}$) \\
\mr
1T & 3.71 & 6.42 & 3.71 & 3.71 & 3.71 & & 2.67  \\
1T$^{\prime}$ & 3.71 & 6.38 & 3.71 & 3.47 & 3.79 & & 2.61  \\
CDW & 6.42 & 6.42 & 3.73 & 3.47 & 3.47 & & 2.64  \\
AABB & 3.71 & 12.27 & 3.71 & 3.79 & 3.39 & & 2.59  \\
1T$^{\prime\prime}$ & 3.73 & 6.07 & 3.73 & 3.47 & 3.47 & & 2.58  \\
Zig-zag & 7.20 & 6.24 & 3.60 & 3.51 & 3.69 & & 2.56 \\
\br
\end{tabular*}
\end{table}

\subsection{Magnetic interactions} \label{sec2}
We now turn to the analysis of the magnetic properties using the all-electron KKR-GF method.
The calculated magnetic interactions are long-ranged and display an oscillatory behavior as function of distance, as expected on metals, which can lead to energetic competition between different magnetic states and to the stabilization of magnetic spirals.
Numerically, we found that the identified ground state is robust once interactions up to a distance of 6 times the nearest-neighbor distance are incorporated in the simulations.
Taking the 1T-phase as an example, if we include only the interactions up to the fourth nearest-neighbor distance we find that the ground state is the non-collinear N\'eel-AFM state, while taking more interactions into account transforms the ground state into a spin spiral state. 
The set of Heisenberg exchange and DM interactions up to fourth nearest-neighbors and the magnetic anisotropy are collected in Table \ref{table:mag-inter}.

The magnetic anisotropy ranging from  \SI{1.2}{\milli\electronvolt} to \SI{1.4}{\milli\electronvolt} favors an in-plane orientation of the magnetic moments in agreement with both experimental \cite{fm-crte2} and theoretical \cite{cdw-crte2} works conducted on the 1T-phase.
In all structures, the magnetic anisotropy is uniaxial, except for the zig-zag phase which has a small additional in-plane anisotropy around \SI{0.3}{\milli\electronvolt}.
One can see that except for the 1T$^{\prime}$ and CDW phases all structures have an AFM first nearest neighbours interaction (J$_{1}<0$) followed by an oscillatory behavior.

\begin{table}[htb]
\caption{\label{table:mag-inter}The Heisenberg exchange coupling ($J$), the magnitude of the Dzyaloshinskii-Moriya vector ($|\vec{D}|$), and  magnetic anisotropy energy ($K$) for the different structures and magnetic phases of the CrTe$_2$ monolayer.
For the DM interaction, --- means that the interaction is zero due to inversion symmetry.
The correspondence to the bonds in each structure is given in Fig.\ \ref{dmi}.}
\begin{tabular*}{\textwidth}{l @{\extracolsep{\fill}} llllllllll}
\br
&&\centre{4}{$J$ (meV)} 
&&\centre{4}{$|\vec{D}|$ (meV)} \\ \cmidrule{3-6} \cmidrule{8-11}
Phase &$K$ (meV)&$J_1$ & $J_2$ & $J_3$ & $J_4$ & & $|\vec{D}_1|$ & $|\vec{D}_2|$ & $|\vec{D}_3|$ & $|\vec{D}_4| $ \\
\mr
1T & 1.4 &-5.4 & 4.0 & 2.0 & -0.9 & & ---  & --- & ---  & --- \\
1T$^{\prime}$ &1.3 & 0.6  & -2.8 & 4.4 &  2.9 & & --- & 0.7 & 0.0 & 0.2\\
CDW & 1.3 & 5.7 & 2.7 & -0.4 & 0.7 & & 1.5  & 0.2& 0.2& 0.3 \\
AABB & 1.2  &-13.6 & -4.9 & -2.4 & 5.1 & & ---  & 0.1 & --- & 0.5\\
1T$^{\prime\prime}$ &1.2  &-13.0 & -0.6 & 1.2 & 1.7 &  & ---  & --- & --- & --- \\ 
Zig-zag &1.2  &-13.4 & -14.9 & -7.6 & 2.1 &  & --- & 0.2 & --- & 0.3 \\ 
\br
\end{tabular*}
\end{table}

\begin{figure}[htb]
     \begin{subfigure}[b]{0.333\textwidth}
        \captionsetup{justification=centering}
        \caption{1T}
         \includegraphics[width=\textwidth]{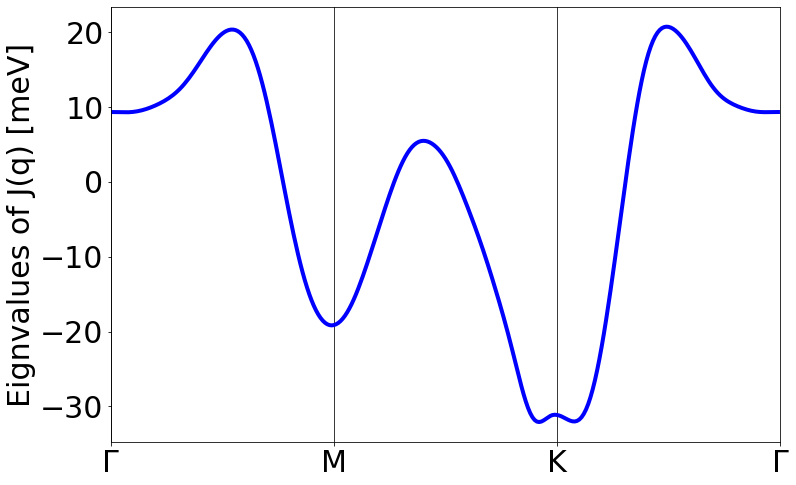}
         \label{1t}
     \end{subfigure}
     \begin{subfigure}[b]{0.333\textwidth}
        \captionsetup{justification=centering}
         \caption{1T$^{\prime}$}
         \includegraphics[width=\textwidth]{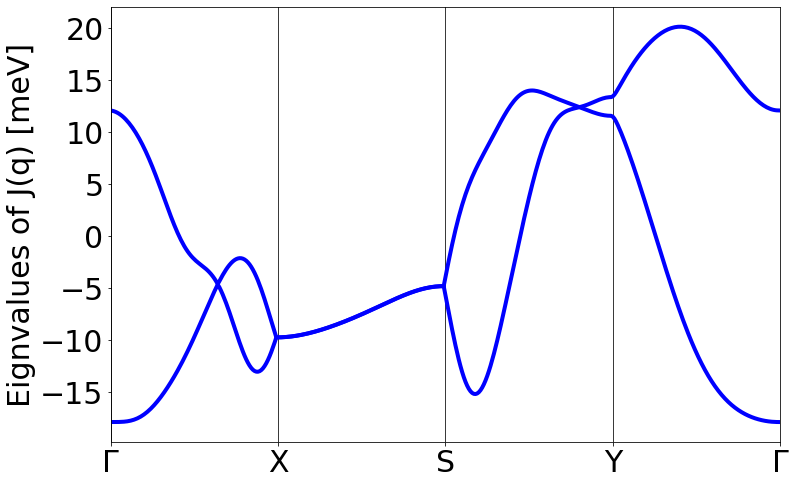}
         \label{1t'}
     \end{subfigure}
     \begin{subfigure}[b]{0.333\textwidth}
        \captionsetup{justification=centering}
         \caption{CDW}
         \includegraphics[width=\textwidth]{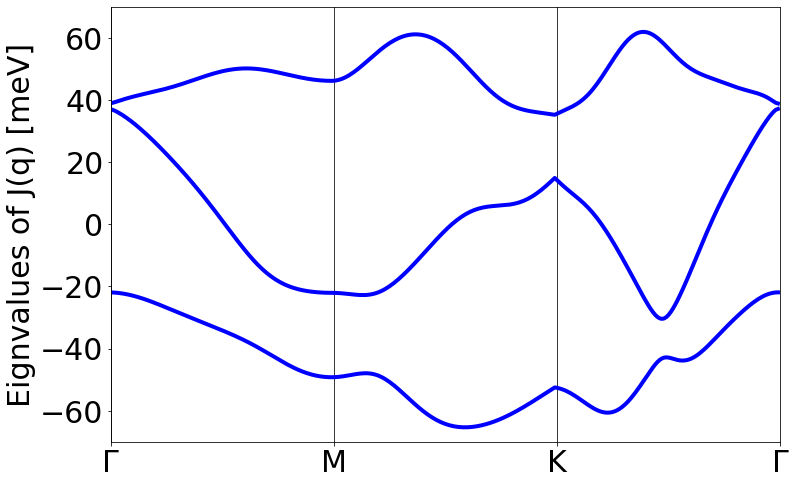}
         \label{cdw}
     \end{subfigure}\vspace{-1em}
     \begin{subfigure}[b]{0.333\textwidth}
         \captionsetup{justification=centering}
          \caption{1T$^{\prime\prime}$}
         \includegraphics[width=\textwidth]{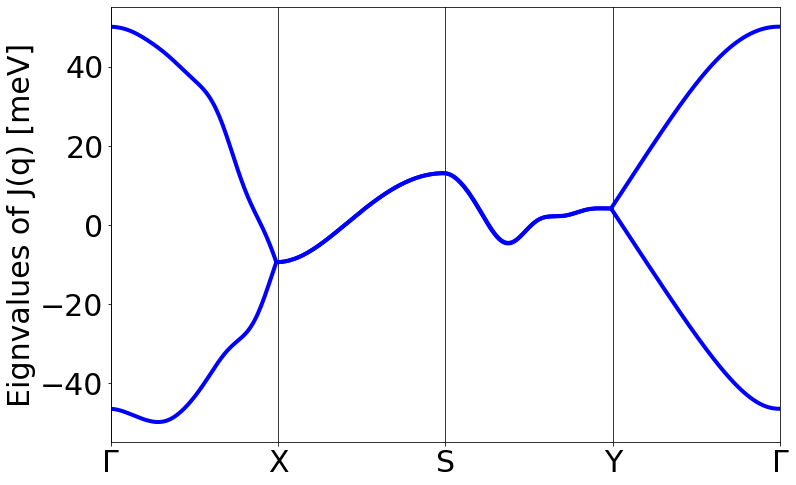}
         \label{1t''}
     \end{subfigure}
     \begin{subfigure}[b]{0.333\textwidth}
        \captionsetup{justification=centering}
         \caption{AABB}
         \includegraphics[width=\textwidth]{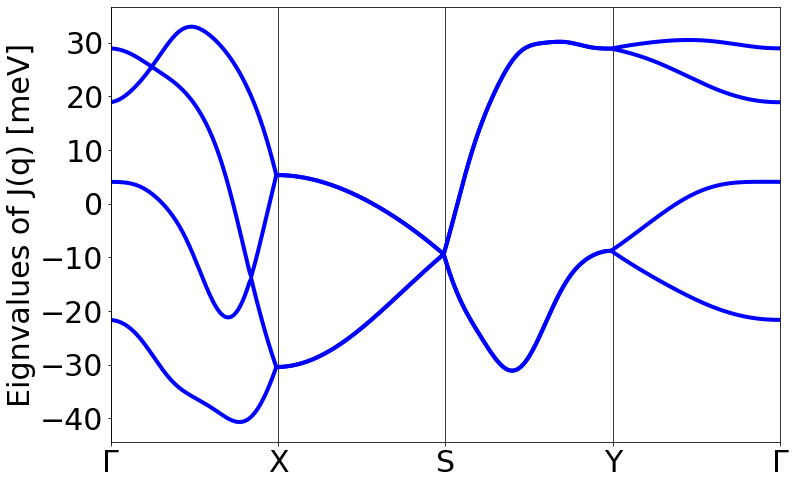}
         \label{aabb}
     \end{subfigure}
     \begin{subfigure}[b]{0.333\textwidth}
        \captionsetup{justification=centering}
         \caption{Zig-zag}
         \includegraphics[width=\textwidth]{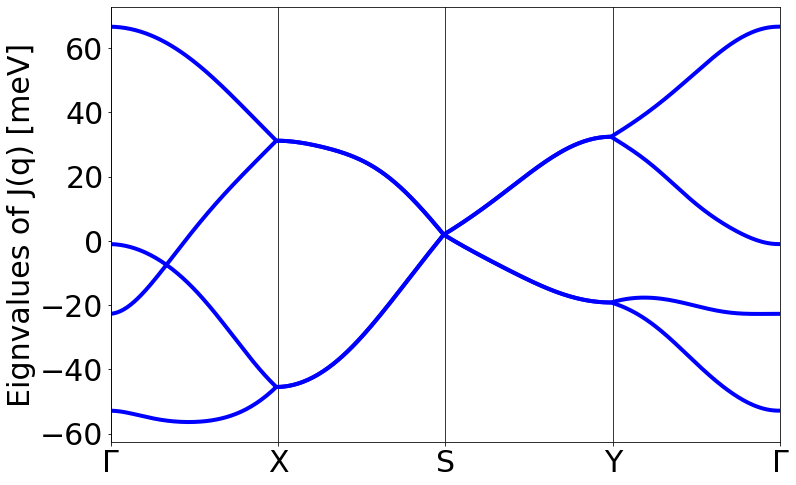}
         \label{zigzag}
     \end{subfigure}\vspace{-1em}
     \begin{subfigure}[b]{0.9\textwidth}
        \captionsetup{justification=centering}
        \caption{}
        \includegraphics[width=\textwidth]{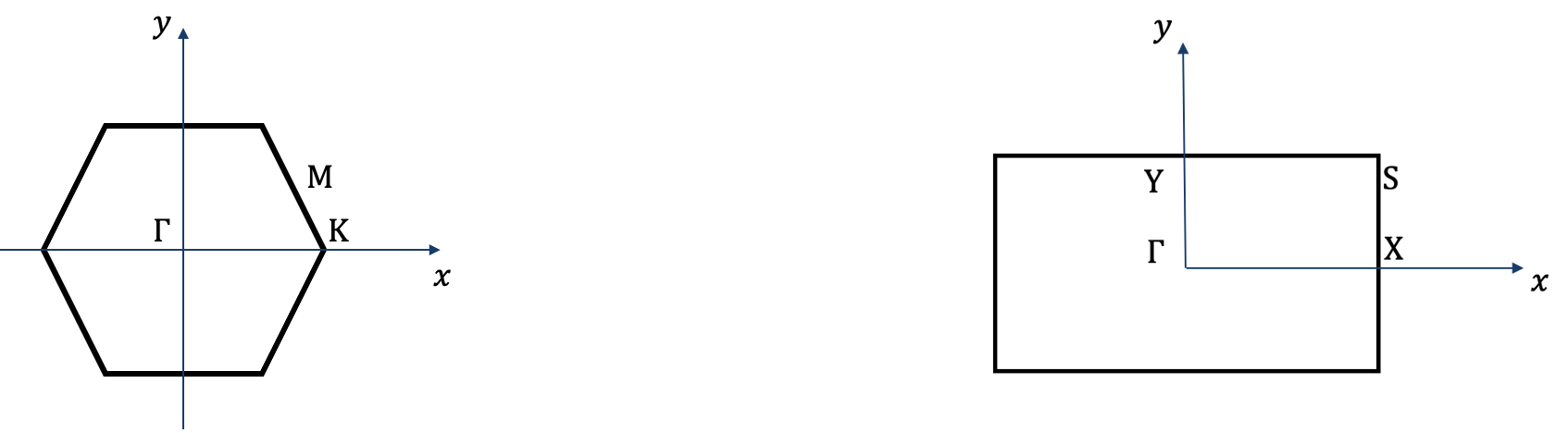}
        \label{fbz-1}
    \end{subfigure}\vspace{-2em}
    \caption{\label{j-q}
        Energetics of magnetic states for the different crystal structures of a CrTe$_2$ monolayer based on the computed exchange interactions.
        (a-f) Eigenvalues of the Fourier-transformed exchange interactions as a function of $q$.
        (g) The hexagonal first Brillouin zone (left) for the 1T and CDW structures, and the rectangular Brillouin zone (right) for the 1T$^{\prime}$,1T$^{\prime\prime}$, AABB, zig-zag structures.}
\end{figure}

In Fig.\ \ref{j-q} we plot the eigenvalues of the Fourier-transformed Heisenberg exchange interactions as a function of reciprocal momentum vector $q = 2\pi/\lambda$ using Eq. \ref{eq:exc_contr}, which serve as proxies for the energy of the magnetic state with the same periodicity.
We see that the 1T phase is characterized by energy minima located near the K-point in the first Brillouin zone, which indicates that the ground state is a spiral state that can form along the M-K or K-$\Gamma$ directions with a wavelength $\lambda \approx 15d_1$, where $d_1$ is nearest-neighbor distance.
The 1T$^{\prime}$ is ferromagnetic (energy minimum at $\Gamma$) while all the AFM structures (1T$^{\prime\prime}$, AABB, and zig-zag) host spiral states as ground states (with wavelengths of approximately $8d_1$, $5d_1$, and $8d_1$, respectively). 

\begin{figure}[htb]
     \begin{subfigure}[b]{0.25\textwidth}
      \captionsetup{justification=centering}
       \caption{1T}
         \includegraphics[width=\textwidth]{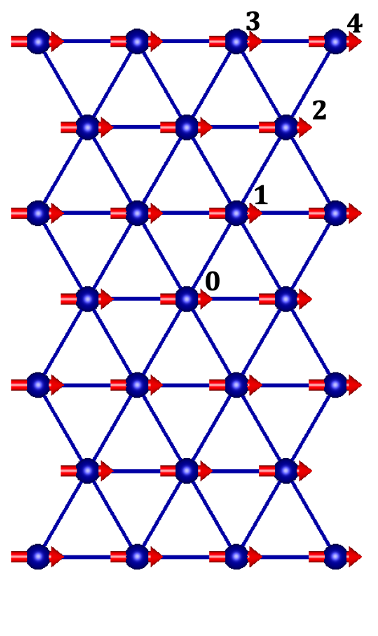}
     \end{subfigure}
    \hspace{0.05\textwidth}
     \begin{subfigure}[b]{0.23\textwidth}
        \captionsetup{justification=centering}
       \caption{1T$^{\prime}$}
         \includegraphics[width=\textwidth]{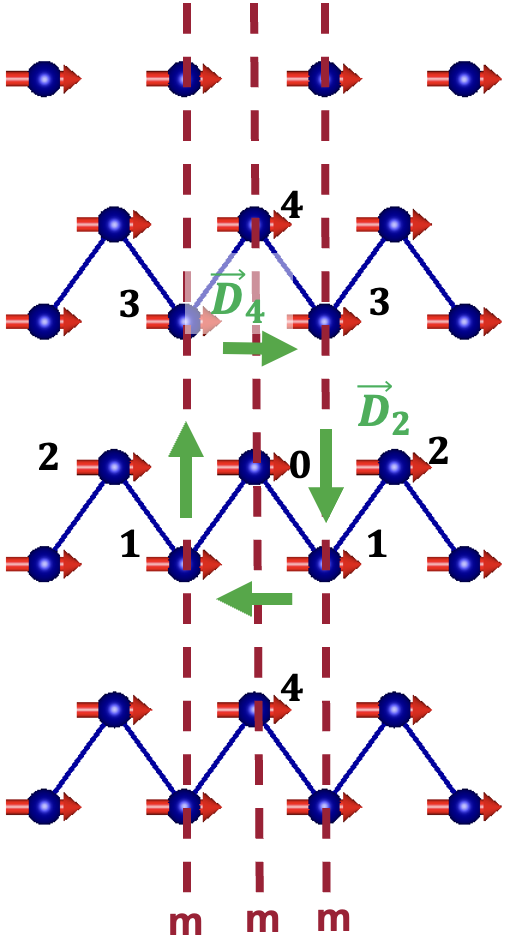}
     \end{subfigure}
    \hspace{0.05\textwidth}
     \begin{subfigure}[b]{0.34\textwidth}
        \captionsetup{justification=centering}
       \caption{CDW}
         \includegraphics[width=\textwidth]{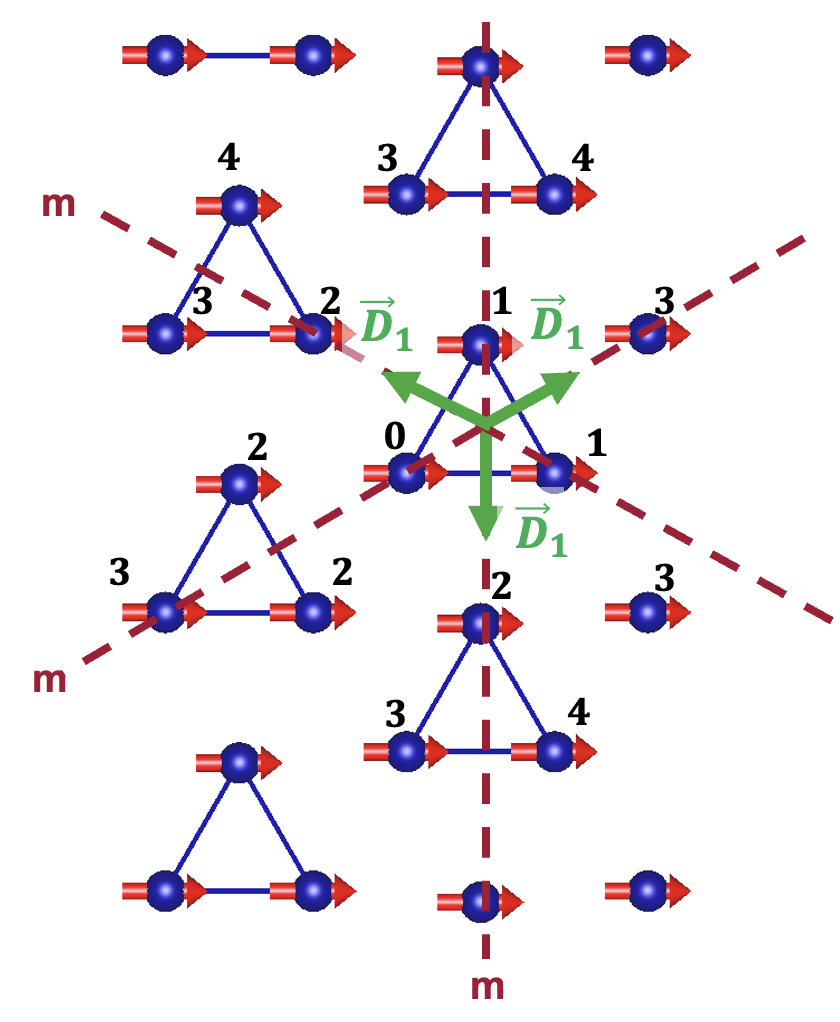}
     \end{subfigure}
    \begin{subfigure}[b]{0.25\textwidth}
        \captionsetup{justification=centering}
       \caption{1T$^{\prime\prime}$}
         \includegraphics[width=\textwidth]{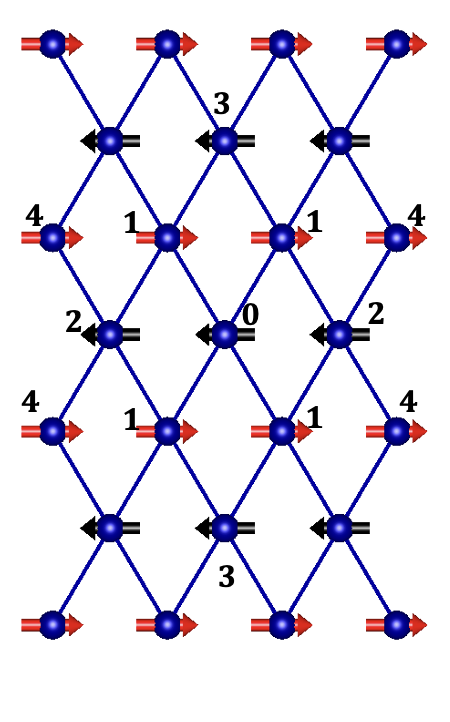}
     \end{subfigure}
     \hspace{0.05\textwidth}
     \begin{subfigure}[b]{0.24\textwidth}
        \captionsetup{justification=centering}
       \caption{AABB}
         \includegraphics[width=\textwidth]{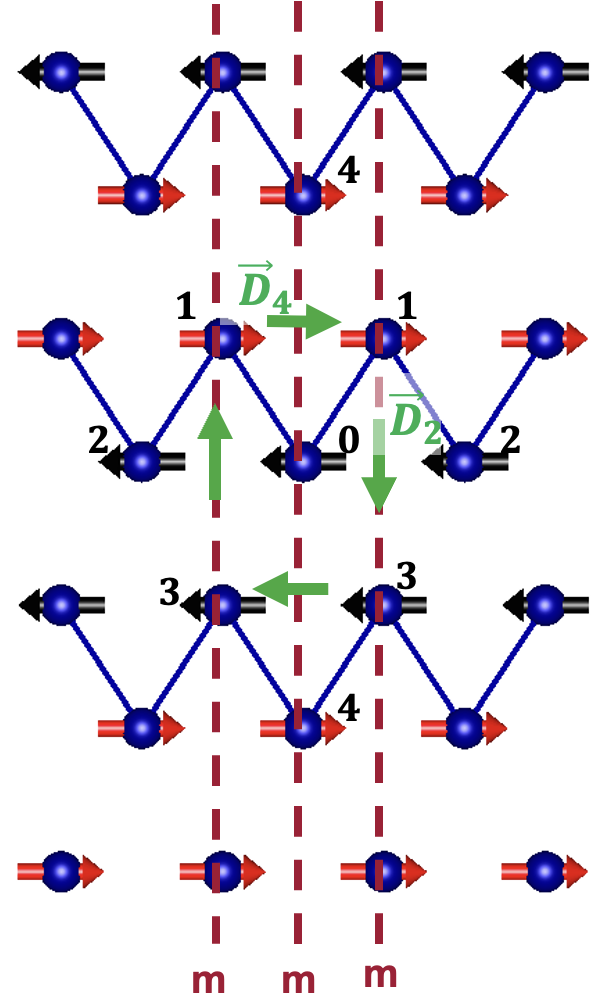}
     \end{subfigure}
     \hspace{0.05\textwidth}
     \begin{subfigure}[b]{0.32\textwidth}
        \captionsetup{justification=centering}
       \caption{Zig-zag}
         \includegraphics[width=\textwidth]{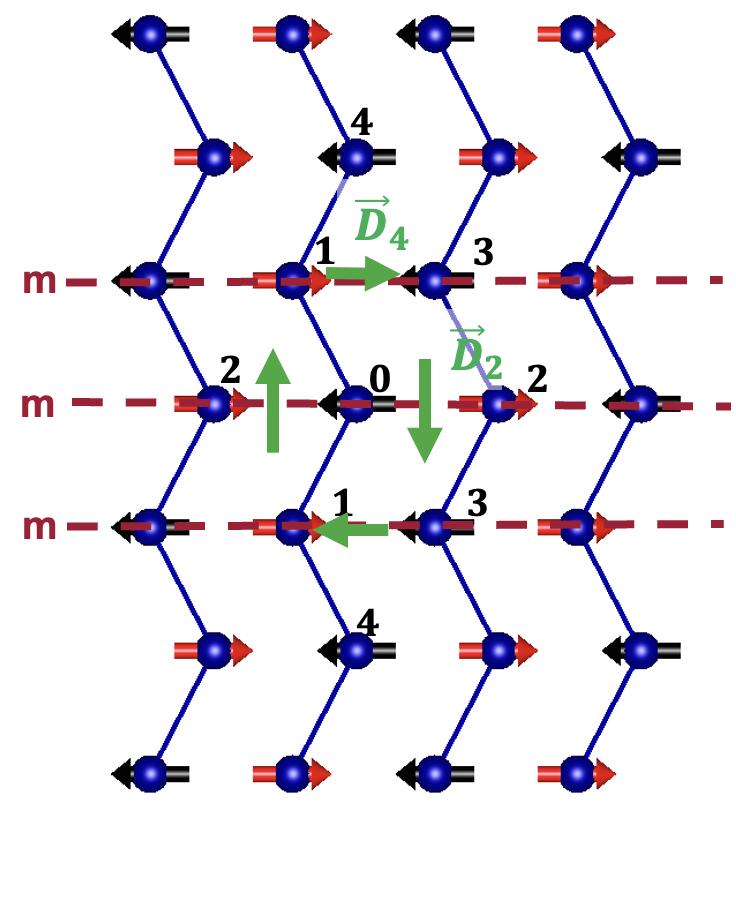}
     \end{subfigure}
    \caption{\label{dmi}
    Dzyaloshinskii-Moriya interactions in CrTe$_2$.
    (a,d) Centrosymmetric structures 1T and 1T$^{\prime\prime}$ for which the DMI vanishes.
    (b-c,e-f) Non-centrosymmetric structures 1T$^{\prime}$, CDW, AABB, and zig-zag for which the DMI is allowed.
    The DM vectors are represented by green arrows, and the mirror planes that enforce the corresponding Moriya symmetry rules by dashed lines.
    The numbers 1 to 4 represent the four nearest neighbors for the reference atom 0, according to increasing bond length.}     
\end{figure}

We now turn to the discussion of the DMI, which is induced by the combination of the spin-orbit interaction and lack of inversion symmetry.
From our structures, those that lack inversion symmetry are the 1T$^{\prime}$, CDW, AABB, and zig-zag, with the magnitude of the DM vector for different pairs listed in Table \ref{table:mag-inter}.
The basic properties of the DM vectors follow from the symmetry rules derived by Moriya \cite{Moriya}.
As shown in Fig.\ \ref{dmi}, the 1T$^{\prime}$ structure has an inversion symmetry point between the pairs 0-1, and 0-3, so that the corresponding DM vector vanishes.
This does not apply to the other two pairs, so $|\vec{D}_2|$ and $|\vec{D}_4|$ are finite. 
Following another of Moriya’s symmetry rules, since the mirror planes (m) are perpendicular to the middle of the bonds between 0-2 pairs, their respective DM vectors lie within the mirror planes. 
However, the same mirror planes pass through the bonds between 0-4 pairs, so the DMI vector for these pairs is perpendicular to the corresponding mirror plane. 
Regarding the CDW structure, the DMI is finite for all nearest neighbors. 
According to the previously mentioned Moriya’s symmetry rules, the DMI vector for the first nearest neighbors lies in the mirror plane perpendicular to the middle of the 0-1 bond.  
The DMI emerging in the AABB structure follows the same rules as in 1T$^{\prime}$ with higher values for $|\vec{D}_2|$ and $|\vec{D}_4|$.
In the zig-zag structure, the the mirror planes are parallel to the middle of the bonds between 0-2 pairs, their respective DM vectors perpendicular to the mirror planes.
However, the same mirror planes perpendicular to the bonds between 0-4 pairs, so the DMI vector for these pairs pass through the mirror planes.

\subsection{Atomistic spin dynamics and magnetic ground states} 
In the previous section, we predicted a spiral state in all structures of the monolayer except 1T$^{\prime}$ which has a collinear FM state as the ground state. 
This spiralization of the unveiled magnetic states is governed by the frustration of the Heisenberg exchange interactions. 
Next we perform atomistic spin dynamics to explicitly visualize the predicted magnetic states and to explore the effect of the two other interactions, the DMI and the magnetic anisotropy. 

\begin{figure}[tb]
    \begin{tabular*}{\textwidth}{c @{\extracolsep{\fill}} ccc}
        (a) &  & (b) &  \\
        & \includegraphics[width=0.4\textwidth]{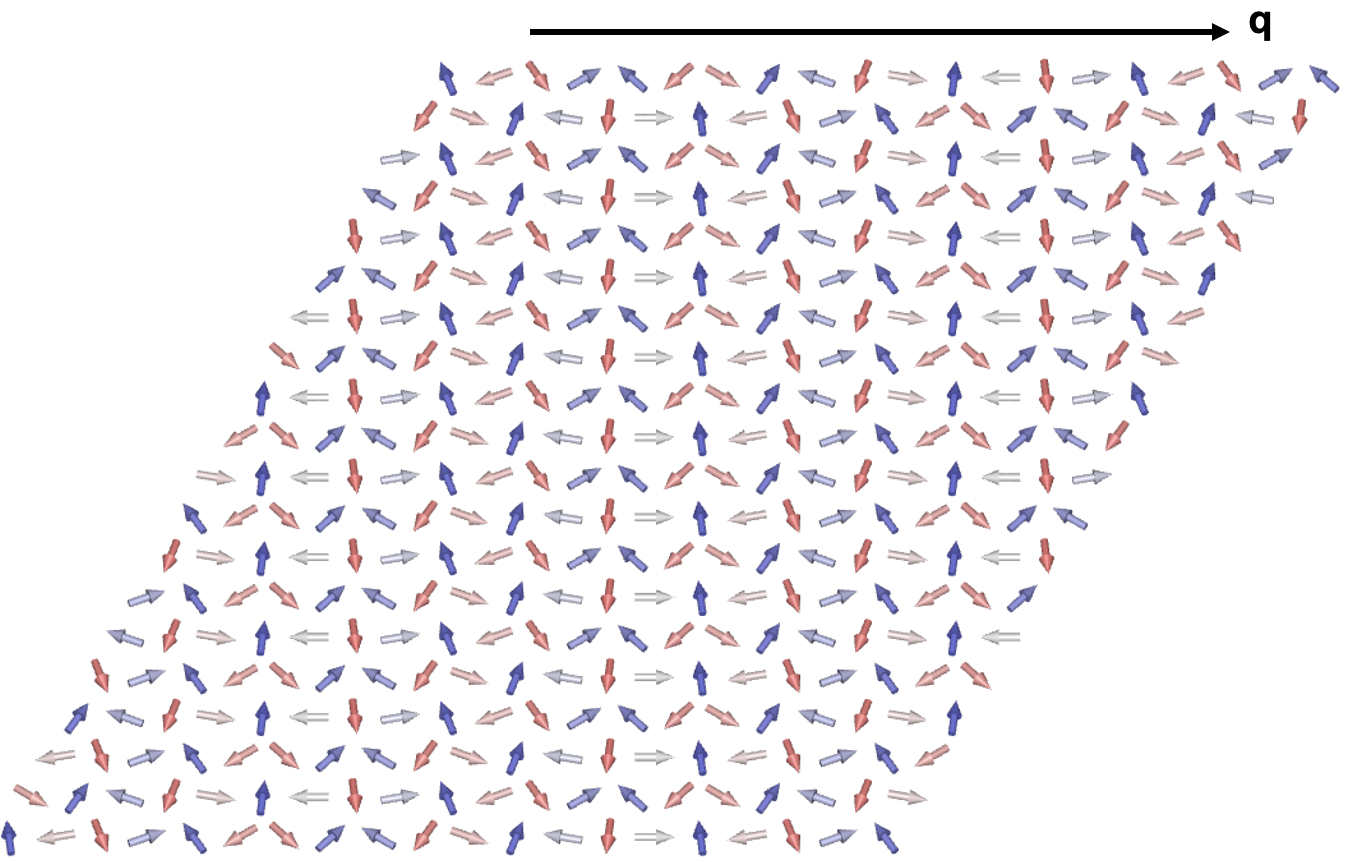} & & \includegraphics[width=0.4\textwidth]{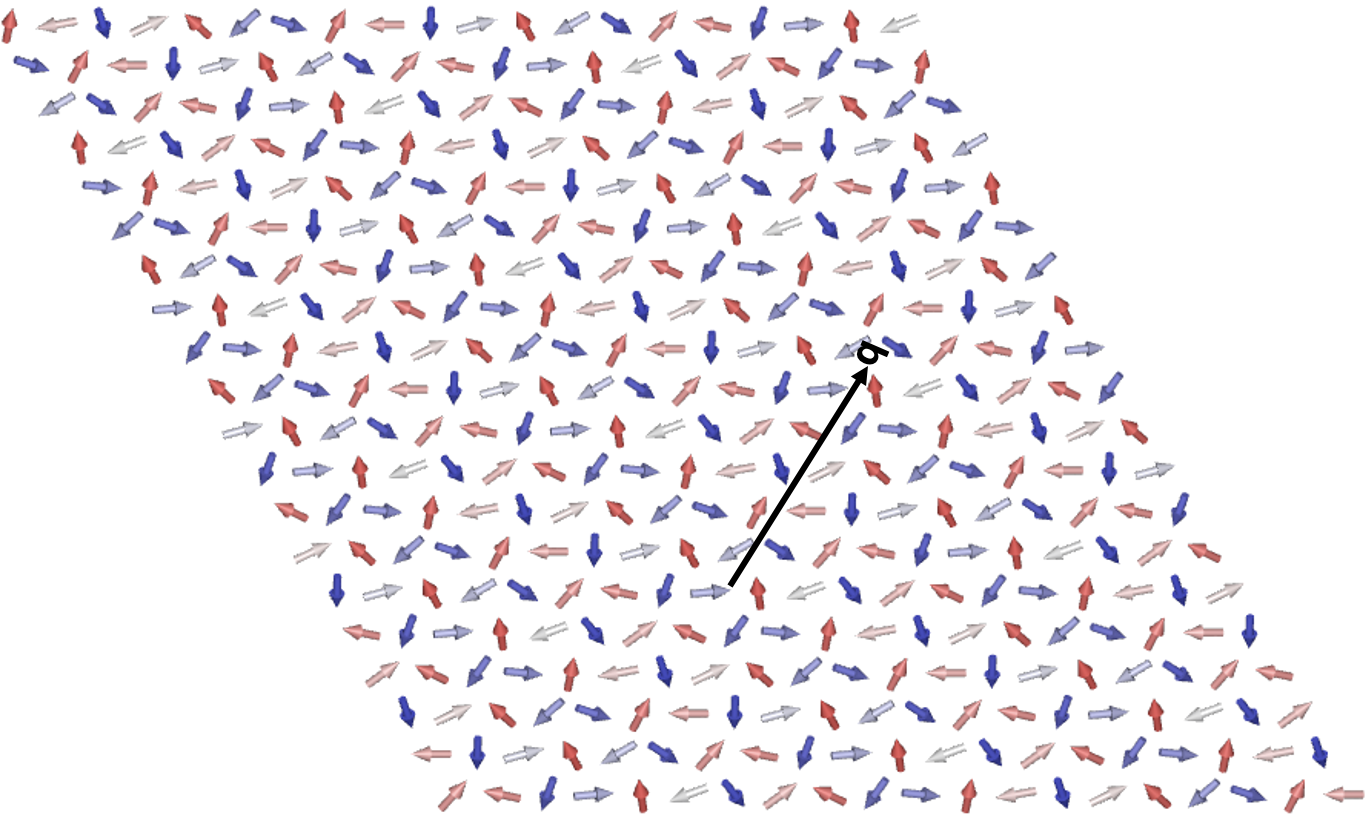} \\
        (c) &  & (d) &  \\
        & \includegraphics[width=0.4\textwidth]{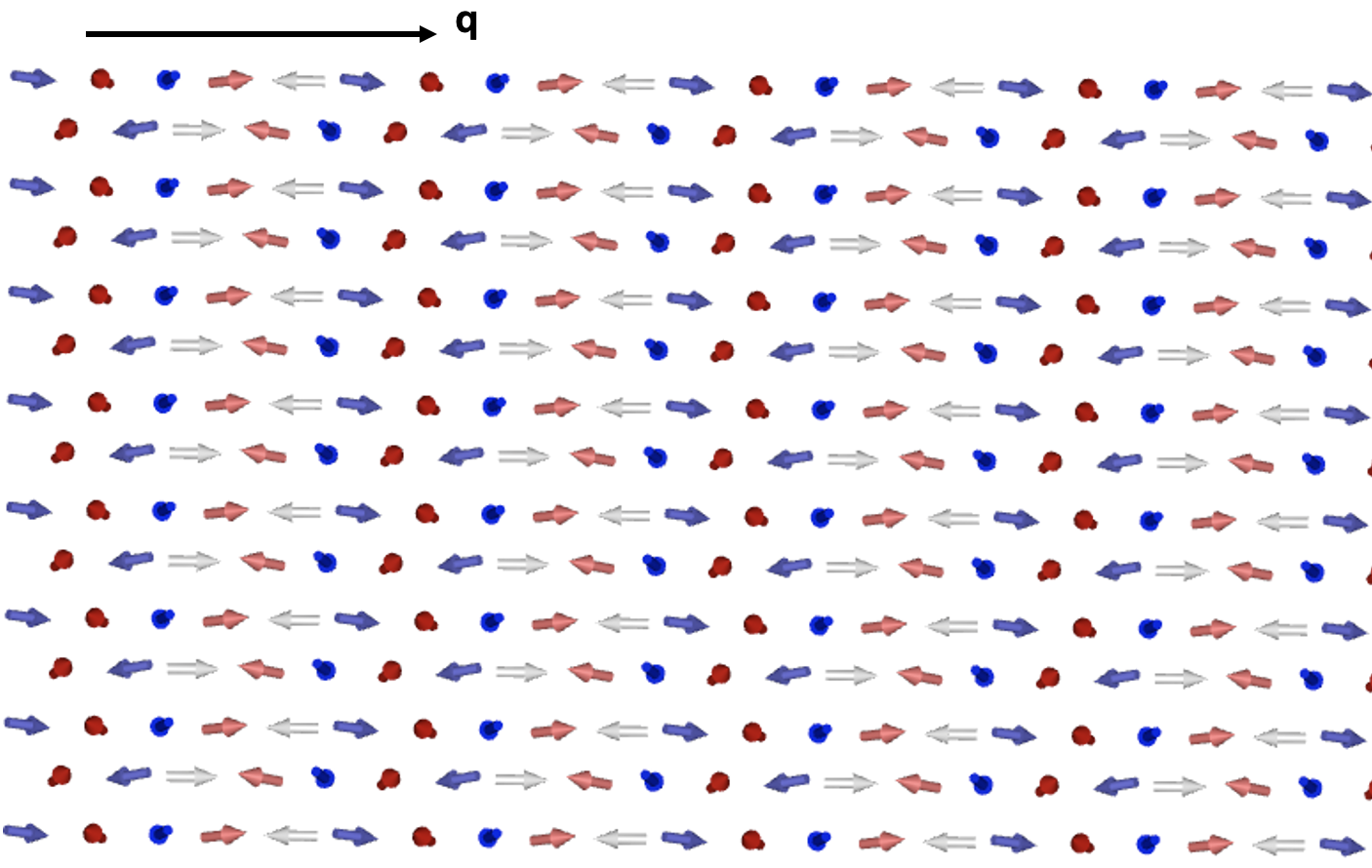} & & \includegraphics[width=0.4\textwidth]{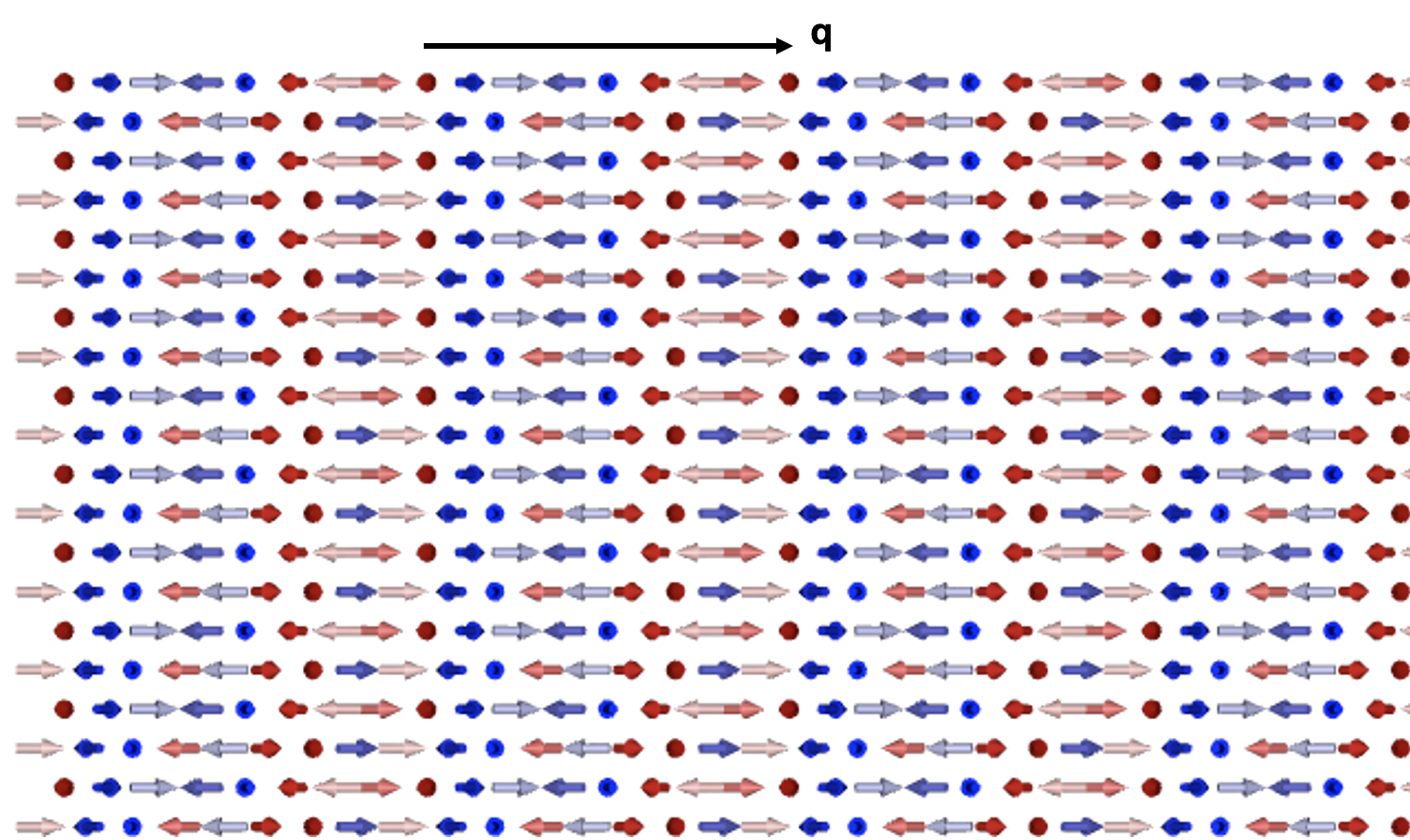}
    \end{tabular*}
    \caption{\label{fig:spindynamics}
    Spiral states in different structures of a CrTe$_2$ monolayer.
    (a) 1T ($\lambda = 15d_1$); (b) CDW ($\lambda = 5d_1$); (c) AABB ($\lambda = 5d_1$); (d) Zig-zag ($\lambda = 8d_1$).
    The 1T$^{\prime\prime}$ case is similar to the AABB but with $\lambda = 8d_1$ and is not shown.}
\end{figure}

The results of these simulations are depicted in Fig.\ \ref{fig:spindynamics}, which reveals that the ground states are qualitatively similar to what was anticipated based on the Heisenberg exchange interactions alone, with the same periodicities as found in section \ref{sec2}.
The impact of the DMI and magnetic anisotropy energy is as follows.
As the magnetic anisotropy is of easy-plane type, so its inclusion in the simulations favors the spins to rotate in the plane of the monolayer.
For the structures in which it is allowed, the DMI favors a specific sense of rotation of the magnetic moments in a plane perpendicular to the DM vectors.
As this plane is typically different of the one favored by the magnetic anisotropy, the two interactions compete against each other.
For the 1T$^{\prime}$ structure, the DMI would stabilize a long-period spin spiral state with wavelength $\lambda \approx 100d_1$; however this is energetically unfavorable when the magnetic anisotropy is accounted for, and so the ferromagnetic state remains the ground state, with an in-plane orientation of the spin moments.
The competition between the DMI and the magnetic anisotropy is exemplified for the case of the AABB structure.
This causes the spins to tilt away from the $xy$-plane favored by the anisotropy, while leaving the periodicity of the spiral state derived from the isotropic exchange interactions essentially unchanged.

Finally, we compare the total energies of each structure to that of the 1T phase (see Table \ref{table:energies}).
We consider two contributions: the total energy differences for collinear reference magnetic states, as calculated with the KKR-GF method (DFT), and the additional energy lowering from each reference magnetic state to the lowest energy state found by atomistic spin dynamics using the Spirit code. Overall, the ground state is found to be the non-collinear AFM zig-zag structure.

\begin{table}[htb]
\caption{\label{table:energies} Combined magnetic energy lowering for each of the considered crystal structures.
$\Delta E_\mathrm{DFT}$ represents the DFT total energy differences between the collinear magnetic states, while $\Delta E_\mathrm{Spirit}$ represents the additional energy lowering found for the non-collinear magnetic states as evaluated via Spirit (spin model).
$\Delta E_\mathrm{total}$ gives the combined energy difference ($\Delta E_\mathrm{DFT} + \Delta E_\mathrm{spirit}$).}
\setlength{\tabcolsep}{26pt}
\begin{tabular}{@{}llll}
\br
Phase & $\Delta E_\mathrm{DFT}$ (meV) & $\Delta E_\mathrm{Spirit}$ (meV) & $\Delta E_\mathrm{total}$ (meV) \\
\mr
1T & 0 & -45.7 & 0 \\
1T$^{\prime}$ & -56.7 & -1.3  & -12.3 \\
CDW & -61.2 & -91.9 & -107.4  \\
AABB & -50.5 & -58.3 & -63.1 \\
1T$^{\prime\prime}$& -230.4 & -4.6 & -189.3 \\ 
Zig-zag & -241.7 & -5.6 & -201.3  \\ 
\br
\end{tabular}
\end{table}

\section{Conclusions}
In summary, using a combination of density functional theory calculations and atomistic spin dynamics we demonstrated that the monolayer of CrTe$_{2}$ can host various structural phases with a rich set of magnetic states.
Interestingly, this 2D material hosts strong magneto-elastic coupling phenomena.
Imposing various magnetic states drives structural transitions, from which emerge new crystal structures with different atomic displacements and deformations of the unit cell.
For each of the obtained phases, non-trivial non-collinear magnetic states are obtained, where the physics is driven by long-range competing exchange interactions.
The DMI is present in some of the structures, which can drive chiral magnetism, and so we assessed its influence.
For the free standing CrTe$_{2}$, we did not find a signature of topological magnetic objects such as skyrmions in both cases, ferromagnetic and antiferromagnetic state.
Although the overall ground state is the AFM spin spiral hosted by the zig-zag structure, we foresee the rich potential tunability of both the crystal structure and the magnetic states depending on the substrate on which the single CrTe$_2$ layer is deposited, whether it is strained, or how it is integrated in van der Waals heterostructures or other types of multilayers.
This motivates to pursue the exploration of this intriguing 2D material in these different scenarios.

When constraining our simulations to collinear magnetism, our results are consistent with the DFT calculations reported in Refs.\ \cite{afm-crte2,thick-crte2}, which indicate that a CrTe$_2$ monolayer has a zig-zag AFM ground state.
The CDW structure identified in Ref.\ \cite{cdw-crte2} is also plausible when restricting the calculations to only ferromagnetic states, and could be stabilized under some experimental conditions.
Notably, the antiferromagnetic zig-zag order was experimentally detected in Ref.\ \cite{afm-crte2} using SP-STM, although the structure was reported to be collinear.
We speculate that the measured response to an external magnetic field and the observed SP-STM contrast difference in comparison with the zero field measurements is naturally explained by the noncollinear zig-zag structure that we found, after accounting for expected differences due to the graphene bilayer and capping, which are not accounted for in our calculations.


\ack
This work was supported by the Federal Ministry of Education and Research of Germany in the framework
of the Palestinian-German Science Bridge (BMBF grant number 01DH16027).  We acknowledge funding provided by the Priority Programmes SPP 2244 "2D Materials Physics of van der Waals Heterostructures"  (project LO 1659/7-1) and SPP 2137 “Skyrmionics” (Projects LO 1659/8-1) of the Deutsche Forschungsgemeinschaft (DFG).
We acknowledge the computing time granted by the JARA-HPC Vergabegremium and VSR commission on the supercomputer JURECA at Forschungszentrum Jülich~\cite{jureca} and RWTH Aachen University under project jara0189.

\section{References}
\bibliographystyle{iopart-num}
\bibliography{references}

\end{document}